\documentclass[twocolumn,showpacs,preprintnumbers,amsmath,amssymb]{revtex4}

\usepackage{graphicx}
\usepackage{dcolumn}
\usepackage{bm}
\usepackage{marvosym}

\begin{document}

\preprint{APS/123-QED}

\title{Capillary origami: spontaneous wrapping of a droplet with an elastic sheet}

\author{C. Py$^{1}$, P. Reverdy$^{2}$, L. Doppler$^{1}$, J. Bico$^{1}$, B. Roman$^1$ \email{benoit@pmmh.espci.fr} and
C. N. Baroud$^2$}
\affiliation{
$^1$ Physique et M\'ecanique des Milieux H\'et\'erog\`enes, ESPCI, Paris~6, Paris~7, UMR CNRS 7635, 75231 Paris cedex5, France\\
$^2$ Laboratoire d'Hydrodynamique (LadHyX) and D\'epartement de M\'ecanique, \'Ecole Polytechnique, UMR CNRS 7646, 91128 Palaiseau cedex, France}
\date{\today}

\begin{abstract}

{\it Abstract} The interaction between elasticity and capillarity is used to produce
three dimensional structures, through the wrapping of a liquid
droplet by a planar sheet. The final encapsulated 3D shape is
controlled by tayloring the initial geometry of the flat membrane.
A 2D model shows the evolution of open sheets to closed structures
and predicts a critical length scale below which encapsulation
cannot occur, which is verified experimentally. This {\it
elastocapillary length} is found to depend on the thickness as
$h^{3/2}$, a scaling favorable to miniaturization which suggests a
new way of mass production of 3D micro- or nano-scale objects.

\end{abstract}



\maketitle

Origami and haute couture design both rely on folding and
assembling planar material to create elegant three-dimensional
(3D) shapes whose variety and complexity is governed by the
number, order and orientation of folds. Folding is also a way to
reduce the size of deployable structures in space industry
(satellite pannels or sun sails~\cite{DeFocatiis02}) as well as in
nature (plant leaves folded in a bud~\cite{kobayashi98}). In micro
and nano-fabrication, the folding of planar structures is a
promising approach to build 3D objects, since most
microfabrication technologies produce planar layers through
surface etching~\cite{madou-book}.

While centimeter-sized objects may be folded using well disposed
magnets~\cite{boncheva05}, capillarity is a particularly
relevant mechanism for micro-fabrication~\cite{boncheva03}, as surface forces
dominate over bulk forces at small scales. Indeed, the capillary
attraction and stiction of wet slender structures may induce disastrous
damage in micro and nano-scale
fabrication~\cite{tanaka93,chakrapani04,mastrangelo93,raccurt04}, in
addition to being involved in lung airway closure (neonatal respiratory
distress syndrome)~\cite{heil02}. Capillary interactions have already
been proposed as a way of assembling and orienting rigid objects in 2D
at the surface of water~\cite{bowden99}, or in 3D by capillary
bridges~\cite{syms03}, while further studies have focused on the
capillary induced deflection of elastic
rods~\cite{cohen03,bico04,kim06}. Recently, the multi-folding of a flexible ribbon squeezed in a meniscus has also been described through an elegant experiment~\cite{kornev06}. Here, we address the effect of
capillary forces on a flat elastic membrane and we show how the
spontaneous folding of a flexible sheet around a liquid droplet leads to
a predetermined 3D shape.

Our experiments were conducted using polydimethylsiloxane (PDMS)
membranes. The PDMS (Dow corning Sylgard 184, 10:1 polymer/curing
agent mix) was spin coated at $24^\circ$C on a glass microscope
slide, at rotation rates of 1000-2000~rpm. Once the PDMS was
cured, this resulted in sheets with thickness in the range
80-40~$\mu$m.

\begin{figure}[ht!]
\centering{\includegraphics[width=6.5cm]{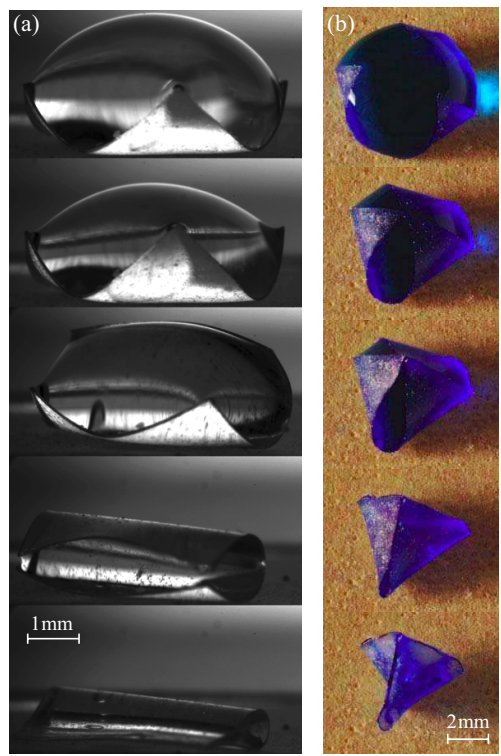}}
\caption{Wrapping of a drop of water with (a) square and (b) triangular
PDMS sheets. The total elapsed time for each experiment is about half
an hour.\label{fig1}}
\end{figure}

During a typical experiment, a geometric shape is manually cut out from
the PDMS layer and placed on a super-hydrophobic surface. A drop of
water, of volume $1-80~\mu$L depending on membrane size, is deposited
on the PDMS, while making sure that the water reaches all the corners of
the flat sheet. The water is then allowed to evaporate at room
temperature and a time lapse sequence of images is taken of the
membrane/drop pair. As the water volume decreases from its initial value
to complete evaporation, the surface tension of the liquid pulls the
sheet around smaller volumes, thus increasingly curving it
(Fig.~\ref{fig1}). Sufficiently thin sheets eventually encapsulate the
liquid with a shape that depends on the initial cut of the membrane.

\begin{figure}[hbp]
\begin{center}
\includegraphics[width=8cm]{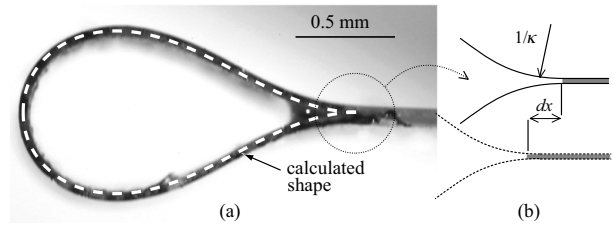}
\caption{(a) Experimental image of a loop of PDMS held by a
meniscus of ethanol, and comparison with the rescaled theoretical
shape (white curve). (b) Virtual displacement $dx$ of the meniscus
position.\label{fig2}}
\end{center}
\end{figure}

In the simplest case, a drop is deposited on a square PDMS sheet, as
shown in Fig.~\ref{fig1}a. This initially leads to the bending of the
four corners towards the center (mode-4).  As the drop volume decreases
further, this bending mode becomes unstable and the shape rapidly
switches to a mode-2 state; in this state, the corners are attracted to
each other two by two and the flat central region disappears, giving
rise to a quasi-cylindrical shape (middle image in Fig~\ref{fig1}a). As
the drop volume decreases further, the approaching edges eventually
touch and a water-filled tube, a few hundred microns in diameter, is
formed. Finally, the water evaporates completely and the tube elarges
into a loop whose ends merge into a flat contacting region, similar to
the shape shown in Fig.~\ref{fig2}a..

A different situation occurs when a drop is deposited on a triangular
sheet, see Fig.~\ref{fig1}b and Movie S1 (www.pmmh.espci.fr/$\sim$benoit/wrapping.avi). The corners fold towards the center as in the
previous case, but this leads to a stable mode-3 folding. The three
corners eventually join at the center and the sheet seals into a
tetrahedral pyramid. At this stage, the curvature of the surface is
concentrated on the edges of the pyramid, while the sides are almost
flat (fourth image in Fig.~\ref{fig1}b). Further evaporation, which now
occurs at a slower rate, finally leads to the buckling of the pyramid
walls due to the negative internal pressure (final image).

For both of the above geometries, the scenario is different for
membranes of smaller size or larger thickness. As the stiffness
increases, the forces generated by surface tension become
insufficient to achieve complete closure: after going through
similar initial stages as for thin PDMS layers, a maximum bending
of the sheet is produced after which it reopens again, never
having produced an encapsulated drop. The final state is therefore
a planar sheet covered by a thin film of liquid which quickly
evaporates.

The membrane deformations reduce the liquid-air area $A$ and thus the
surface energy $\gamma A$ ($\gamma$ being the surface tension), at the
cost of increasing the elastic energy. For a thin plate, the isometric
bending energy density is defined locally as as $B\kappa^2/2$, where
$\kappa$ is the curvature and $B=Eh^3/12(1-\nu^2)$ is the bending
stiffness ($E$ is the Young's modulus, $\nu$ is Poisson's ratio and $h$
is the thickness)~\cite{landau}. For a system of typical dimension $L$,
folding will occur if the surface energy ($\gamma L^2$) is large
compared to the total bending energy ($B$). This is equivalent to
comparing $L$ to

\begin{equation}
L_{EC}=(B/\gamma)^{1/2},
\label{eq lec}
\end{equation}
that we refer to as the {\it elastocapillary
length}~\cite{bico04}. Physically, $L_{EC}$ gives the typical
radius of curvature generated by capillary forces for a given
bending rigidity.

The elastocapillary length can be measured directly through a
calibration experiment rather than by estimating each parameter in
Eq.~\ref{eq lec} which would produce large errors. The experiment
consists in depositing a drop of wetting liquid onto a very long strip
of PDMS, which folds over into a loop with a self-contacting tail. We
consider the late stage, when only a small meniscus still holds the loop
together (Fig.~\ref{fig2}a).

This equilibrium state is 2D, so the following analysis is understood to be per
unit depth. The cross-section follows planar rod elasticity coupled with
surface tension. We note $\theta(s)$ the angle made by the tangent to
the rod $\bf{t}$ with respect to the horizontal, at curvilinear
coordinate $s$. We also note $\bf{R}$ the constant vectorial tension of
the beam, so that Euler's elastica equations for equilibrium may be
expressed as~\cite{landau}

\begin{equation}
B \frac{d^2\theta}{ds^2} \bf{e_z}  + \bf{t} \times \bf{R} =0,
\label{eq elastica}
\end{equation}
where $\bf{e_z}$ is the unit vector perpendicular to the plane of the
section. The global shape of the loop is determined by elasticity,
given the curvature at the contact point. Since this curvature is
set by an equilibrium between surface tension and elasticity, it
must scale as $1/L_{EC}$. The prefactor is found by considering a
virtual displacement $dx$ of the contact point for a loop of curvilinear length  $2\ell$ (see Fig.~\ref{fig2}b). The elastic energy,
\mbox{$E_{el}=\int_{-\ell}^{\ell}B\kappa^2/2\,ds$}, varies due to
the change in the integral bounds, the integrand being stationary
near elastic equilibrium. It may be written as
$dE_{el}=B\kappa_c^2dx$,
where $\kappa_c$ is the curvature at the contact point.
On the other hand, the variation of surface energy due to replacing a
solid-gas interface (surface energy $\gamma_{SG}$) with a
solid-liquid interface ($\gamma_{SL}$) is
$2(\gamma_{SL}-\gamma_{SG})dx=2\gamma\cos\alpha dx$,  where $\alpha$ is the contact angle of the liquid on the surface (here we use ethanol
which has a zero contact angle $\alpha$ on the PDMS). Balancing the two energies yields
the value of equilibrium curvature at the contact point:

\begin{equation}
\kappa_{c} = \frac{\sqrt{2}}{L_{EC}}.
\label{eq condcoll}
\end{equation}

The boundary value problem of Eqs. (\ref{eq elastica}) and (\ref{eq
condcoll}) was solved numerically by enforcing the contact condition at
the base of the loop, $\int_{-\ell}^{\ell} \sin(\theta)ds=0$, and $L_{EC}$ was measured by scaling the numerical solution to fit the
experimental shape. An excellent agreement is obtained (white line
superimposed on experimental picture in fig~\ref{fig2}a). In practice
the width of the loop (theoretical value $0.89 L_{EC}$) is sufficient to
determine $L_{EC}$.

\begin{figure}[htbp]
\begin{center}
\includegraphics[width=7.5cm]{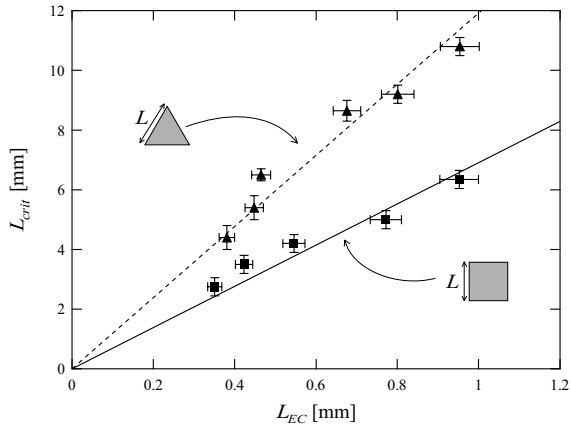}
\caption{Critical length for folding vs. elasto-capillary length: black squares:
square sheets, black triangles: triangular sheets
(leading to 3D pyramids), lines: linear regressions.}
\label{fig3}
\end{center}
\end{figure}

The value of $L_{EC}$ was measured for different membrane thicknesses
and for each thickness, the minimum length for encapsulation $L_{crit}$
was determined for square and triangular membrane shapes. We found that
$L_{crit}$ varied linearly with $L_{EC}$ for both shapes, as one might
expect from dimensional arguments. The slopes were 7.0 for the squares
and 11.9 for the triangles (Fig~\ref{fig3}).

\begin{figure}[htbp]
\begin{center}
\includegraphics[width=7cm]{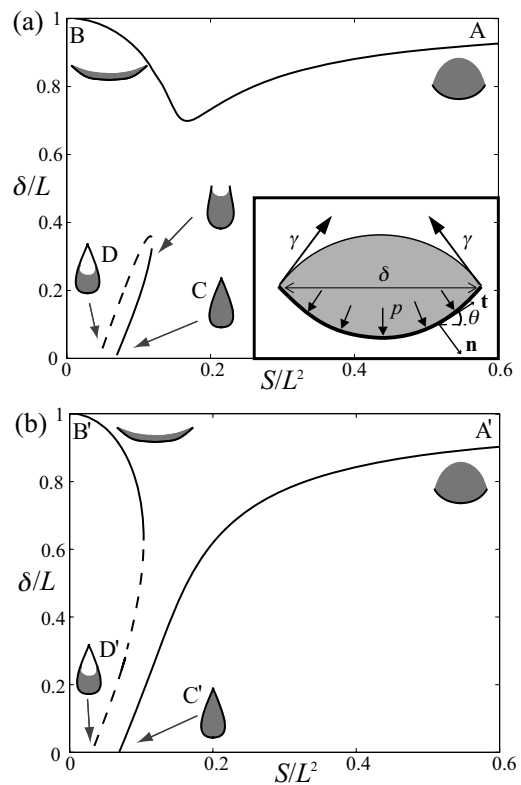}
\caption{Distance between the rod tips versus drop volume for different
values of the rod length $L$: (a) $L =3.4 L_{EC}$ re-opening of the rod as
$S$ decreases; (b) $L =4 L_{EC}$ complete wrapping of
the rod around the drop.  The inset shows a sketch of a 2D drop
deposited on a flexible sheet.\label{fig4}}
\end{center}
\end{figure}

An understanding of the evolution of the folding is obtained from a 2D
theoretical description of the elasto-capillary balance: let us consider
an elastic inextensible rod of length $L$ bent by pressure and surface
tension forces at its ends (Fig.~\ref{fig4} inset). Equation~(\ref{eq
elastica}) still holds but the vectorial tension $\bf{R}$ is not
constant anymore due to the pressure forces but follows

\begin{equation}
\frac{d\bf{R}}{ds} + p \bf{n} =0,
\label{eq pression}
\end{equation}

\noindent where $\bf{n}(s)$ is the unit vector normal to the rod. By
neglecting gravity, the pressure $p$ in the drop is uniform and the 2D
shape of the liquid-air interface is a circular arc of radius
$r=\gamma/p$.

Equations (\ref{eq elastica}) and (\ref{eq pression}), with
appropriate boundary conditions, are solved numerically and the
distance $\delta$ between the rod tips is calculated as a function
of drop section area $S$ for different rod lengths $L$, as shown in
Fig.~\ref{fig4}. Two trivial flat states ($\delta/L=1$) exist in all
cases for an infinitely large drop (A and A') or when no liquid is
present (B or B'). However, the evolution between these states is
qualitatively different depending on $L/L_{EC}$.

For $L/L_{EC}$ slightly below the critical value (Fig.~\ref{fig4}a),
these two states are connected by a continuous family of solutions: the
drying of a large drop leads to transient limited bending but eventually
to reopening. Closed states (C) do exist in a restricted domain of the
phase space, as the stable segment of a solution loop, while the unstable part of this loop corresponds to a drop which has depinned from the edges
(D). This bistability is indeed observed in the experiments: while small
sheets never close spontaneously, they can be forced into a closed state
if the liquid is pulled out with a syringe.

Above the critical ratio $L/L_{EC}$, the drying of a large drop (A') leads
continuously to complete wrapping (C'), as observed experimentally in
Fig.~\ref{fig1}. Another branch links the open state (B') to unstable
solutions (D'). However, this branch is difficult to observe experimentally  because it requires the spreading of a thin film of water on
the flat membrane. Finally, figure~\ref{fig4} shows that the initial
conditions play a major role in selecting the final observed state
because of the presence of bistability.

In the numerical simulations, the crossing from diagrams (a) to (b) was
found at $L_{crit}=3.54~L_{EC}$. This is below the value obtained for
the square membranes, even though the final state of a folded square is
2D (Fig.~\ref{fig1}a). This discrepancy is likely due to gravity and 3D
effects, neither of which is accounted for in the model. Indeed, the
drops considered in Fig.~\ref{fig3} are slightly flattened by gravity,
since they are larger than the capillary length ($L_c=\gamma/\rho g$,
where $\rho$ is the density and $g$ the acceleration of gravity). More
importantly, the early stages are always 3D as seen in the folding of
the corners (e.g. mode-4 folding) and thus deviate the evolution from
the prediction of the 2D model. A 3D theoretical model should include
the full description of thin plate elasticity. Only such a model would
account for the localization of the curvature around the edges of the
pyramid for example, which are reminiscent of the crumpling of thin
sheets.

More generally, 3D wrapping of a droplet leads to a geometrical
incompatibility described by Gauss's {\it theorema egregium}. A
classic example is that it is impossible to draw a planar map of
the earth which conserves distances: the wrapping of a spherical
object with a planar sheet must involve stretching as well as
bending. For a thin membrane, the stretching energy scales as the
thickness $h$ and is therefore much more expensive than the bending
energy which scales as $h^3$. The unavoidable stretching is
therefore localized in singular regions, leading to crumpling
singuliarities~\cite{lobkovsky95,benamar97}. We expect that for
$L\gg L_{EC}$, the number of singularities will increase since
surface tension becomes dominant over elastic forces, leading to
quasi-spherical encapsulated shapes.

\begin{figure}[htbp]
\begin{center}
\includegraphics[width=8cm]{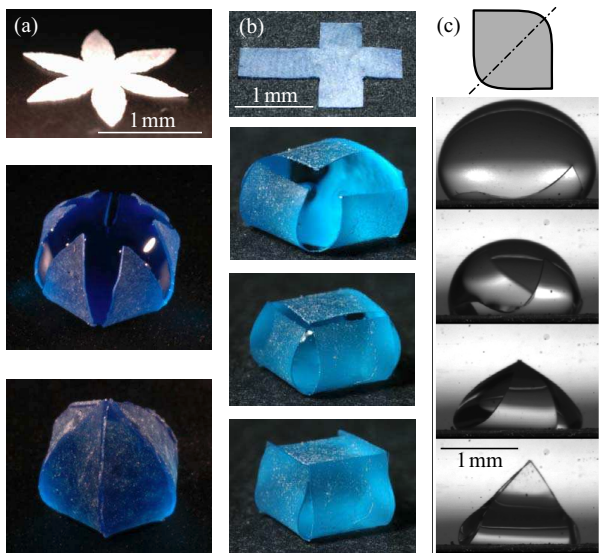}
\caption{Tuning the initial shape to obtain (a) a spherical
encapsulation, (b) a cubic encapsulation or (c) a triangular mode-2 fold. \label{fig5}}
\end{center}
\end{figure}

Precise engineering of the final closed state may be obtained by
tayloring the initial sheet geometry. In this way, it is possible
to approximate a smooth sphere by starting with a flower shape, as
shown in Fig.~\ref{fig5}a. Figure~\ref{fig5}b shows how to form a
cube by starting from a cross shape.
 As in the case of the pyramid,
evaporation after encapsulation takes place at a reduced rate and may
buckle the sphere or the cube due to the negative pressure thus
generated. However, one may also freeze (or polymerize) the encapsulted
liquid at any point in this process, therefore fixing the desired shape
in place. Finally, small perturbations of the initial shape may also be
used to yield different final states. In particular, rounding-off two
opposite corners of a square sheet leads to the membrane closing along
its diagonal (Fig.~\ref{fig5}c) rather than parallel to the sides.
These experiments suggest that a wide variety of final shapes may be
achieved through careful tuning of the initial 2D shape.

In summary, three dimensional sub-millimetric objects are produced from
2D membranes through the interaction of elasticity and capillarity. The
limitation on the minimal size for folding is determined by the
elastocapillary length (Eq.~\ref{eq lec}) which scales as the sheet
thickness $h^{3/2}$. This scaling is favorable to miniaturization since
thinner membranes lead to much smaller critical lengths. This opens the
way to mass production of micron-scale 3D structures based on standard
microfabrication methods since a wide variety of objects can be
fabricated by tayloring the initial planar sheet.


We are very grateful to Emmanuel de Langre (LadHyX, Ecole Polytechnique) for his important
insights. This work was partially supported by the French ministry of
research (ACI {\em Structures \'elastiques minces}), and the
Soci\'et\'e des Amis de l'ESPCI.

\bibliography{origami}

\end{document}